# Towards Blockchain-based Remote Management Systems for Patients with Movement Disorders

Behnaz Behara
*Department of Biomedical Engineering,*
*Faculty of Electrical Engineering,*
*K. N. Toosi University of Technology,*
Tehran, Iran
b.behara@email.kntu.ac.ir

Mehdi Delrobaei
*Department of Mechatronics Engineering,*
*Faculty of Electrical Engineering,*
*K. N. Toosi University of Technology,*
Tehran, Iran
delrobaei@kntu.ac.ir

*Abstract*—Secure storage and sharing of patients' medical data over the Internet are part of the challenges for emerging healthcare systems. The use of blockchain technology in medical Internet of things systems can be considered a safe and novel solution to overcome such challenges. Patients with movement disorders require multi-disciplinary management and must be continuously receive medical care by a specialist. Due to the increasing costs of face-to-face treatment, especially during the pandemic, patients would highly benefit from remote monitoring and management. The proposed work presents a model for blockchain-based remote management systems for patients with movement disorders, especially those with Parkinson's disease. The model ensures a high level of integrity and decreases the security risks of medical data sharing.

## I. Introduction

Due to rapid technological developments, the Internet of things (IoT) significantly makes various parts of human life smarter. The applications of the IoT in multiple fields, such as smart cities, smart agriculture, and smart healthcare systems, have made significant progress. After the increase in healthcare costs and utilization, especially during the pandemic patients' remote management development and improvement seem to be required now more than ever [1].

Therefore, patients' homes and healthcare facilities can gradually be equipped with more IoT-based devices and decision-support algorithms to move faster toward real medical Internet of things technology (IoMT) [2]. Electronic healthcare record (EHR), an innovation introduced along with the emergence of wearable sensors, allows IoT platforms to remotely record patients' information and increase interaction between patients and physicians [3].

The critical challenge in developing remote healthcare management systems is to ensure security. The lack of security not only enables hackers to access personal data but may also lead to damages such as data loss or interruption of service [4]. Therefore, it is essential to develop remote patient management systems while ensuring the security of data transmission and users' privacy [5].

Most of the systems developed for healthcare and remote monitoring of patients use a centralized architecture based on client-server model. One of the significant problems that client-server-based systems face is the lack of privacy and security. This potentially causes cyber attacks and access patient data [6]. Due to the vulnerabilities of client-server systems to cyber-attacks and hacker penetrations, a blockchain-based system can be used for remote patient management and data storage.

Blockchain is a distributed ledger technology that offers a transparent, shared, and smart ledger. In recent years, research related to blockchain has gained significant importance [7], [8]. Blockchain organizes user transactions in the form of a block. Therefore, each block contains several transactions. These blocks are broadcasted to a peer-to-peer chain network, and each block is appended to the previous block so that each block stores the encrypted hash of the last block. Each peer in the blockchain can access a copy of the distributed ledger because the ledger is shared among all peers in the blockchain network [9].

Meanwhile, blockchain technology is used not only in cryptocurrency but also to implement various scenarios and fields, such as supply chains, multimedia, artificial intelligence, cloud computing, and healthcare systems [10], [11]. Among the many applications of blockchain, the use of blockchain technology in health care has received much attention due to its capabilities in medical data security [12]. So far, many blockchain-based IoT healthcare monitoring systems have been designed and developed by various researchers [13]- [15].

Healthcare systems based on the IoT significantly monitor patients who need medical care in real-time. Patients with movement disorders, especially those with Parkinson's disease (PD), would highly benefit from remote monitoring systems. This article generally introduces a remote monitoring system based on the IoT and blockchain for patients with movement disorders.

It is worth noting that many systems have been built to monitor PD patients remotely, but the main limitation is that none guarantee data security over the Internet. Therefore, in order to maintain the security and integrity of data and their safe storage, the idea of blockchain was employed in this work to design such a system.

## II. RELATED WORK

Various blockchain-based systems have been de- signed and developed in the healthcare sector [16]- [18]. Many studies have shown that the use of blockchain in the design of electronic healthcare record systems increases the security and integrity of patient data. In [19], researchers have used blockchain technology and privacy- preserving techniques to maintain the privacy and security of data shared on an IoT platform.

Yanez et al. in [20] proposed a new context-aware data allocation method and designed a controller based on fuzzy logic. They calculated each data request allocation rating (RoA) value by considering parameters such as data and network and deciding its on-chain allocation and quality. The results of this mechanism have shown that using the proposed approach has increased network usage, latency, and blockchain storage and, thus, reduced energy consumption.

In [21], Srivastava et al. presented a blockchain-based monitoring system for remote patient monitoring in the IoT platform. To ensure data security, encryption technologies have been used in the proposed IoT-based blockchain system. In [22], Novo et al. presented a new architecture for blockchain-based IoT systems that arbitrate roles and permissions in IoT. The results of this research indicate that blockchain can be used as an access management technology in various IoT scenarios. Jiaren Cai et al. used SSL and DSL protocols to authenticate and share healthcare data [23].

Liang et al. have proposed a blockchain-based system called Medshare to increase data security in the design of healthcare systems. In this work, blockchain manages access control, provenance, and security of medical records within cloud storage [24]. A secure healthcare electronic system was presented in [25] using blockchain technology and an attribute-based crypto-system approach. The proposed research employs a blockchain-based architecture to develop EHR systems to protect patients' data and medical records against tamper and abuse by others [26]. In [27], researchers have used an end-to-end architecture designed based on a patient-centered agent (PCA) for continuous monitoring of patients. The simulation results indicate that the system increased the security and privacy of patients.

## III. PROPOSED ARCHITECTURE

This section represents a layered architecture for the proposed model. The model suggests a secure management system for remote monitoring of patients with movement disorders, especially PD patients. The layers of the proposed architecture are shown in Fig. 1. The key layers of the proposed architecture include: (1) the user layer, (2) the data processing layer, and (3) the data storage layer. The user layer is where the users interact with the system.

The users include patients, caregivers, nurses, physicians, and IoT devices. The IoT devices collect the Parkinsons patients motion capture data and store them in the form of files. The files will then be uploaded to a remote server. The second layer is the core of the proposed system, implemented on the Ethereum blockchain. The blockchain ensures the security of the users' interactions with the remote monitoring system. The third layer, the storage layer, mainly includes the required smart contracts and the interplanetary file system storage (IPFS).

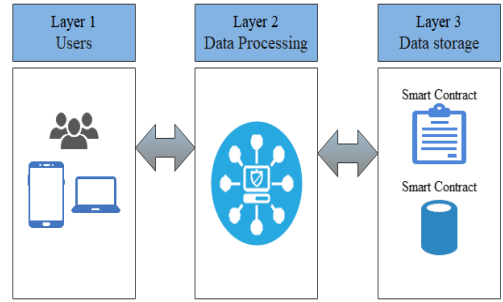

Fig. 1. layers of the proposed architecture.

- **Smart contract:** Smart contracts are used to implement and monitor user tasks and implement various system operators. It assesses the authentication of users who are supposed to interact with this system, and in this case, it can cause secure data transfer and increase the system's safety against unauthorized users. Smart contracts are responsible for checking the data integration, data transfer, record creation, data sharing, data access level control, and access level creation for different users.
- **Interplanetary file System (IPFS) storage:** The IPFS is a peer-to-peer network to store and shares healthcare data. The data collected from users and the IoT devices are encrypted by the Advanced Encryption Standard (AES) algorithm and then uploaded to the IPFS network. Unlike the client-server model, IPFS uses a unique hash. When an encrypted file is stored in IPFS, a unique hash is created from each file to ensure secure data transfer, and the unique hash is stored on the blockchain network. Therefore, IPFS is an ideal and suitable storage system for the secure storage of healthcare data and the medical history of patients.

According to Fig. 2, the proposed model for remote management of patients with movement disorders consists of the following components:
- Smart Contract
- Interplanetary file system (IPFS) storage
- Administrator
- Users
- Decision support algorithm

The administrator (admin) is one of the users of the system. Admin is generally responsible for authenticating and confirming users' registration. The users of the system include:

*a) Patient with movement disorders:* The patient sends their request for registration in the system along with their details, such as name, surname, national number, and medical records, to the administrator through a smart contract. After registering and confirming the patient's information, the administrator sends the patient's login permission to the system along with the user and password. Therefore, the patient can log in to the system and upload or download his medical file and prescriptions. Before uploading the file, the AES algorithm encrypts the data to maintain data security and then stores it in the IPFS.

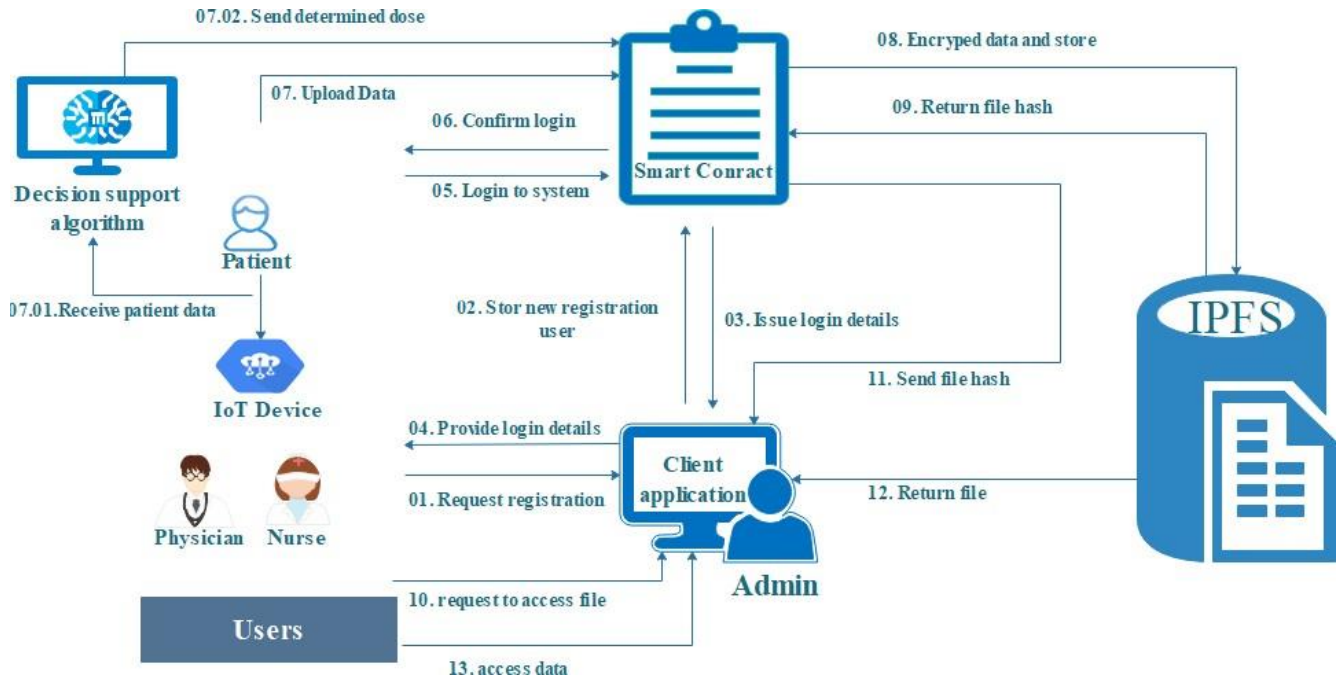

Fig. 2. The proposed model for remote management of patients with movement disorders.

Finally, a unique hash file is given to the patient. In fact, with this hash file, the patient can share data with the physician. By logging into the system and user panel, the patient can choose a physician from the list of available physicians and send his unique hash file to the physician. It is noteworthy that due to the use of the blockchain platform, all the patient's information and history will be recorded as immutable in the system.

*b) Physicians:* The process of registering and logging into the system for physicians is similar to registering patients. After logging into the system, the physician can view the list of patients who have sent their requests. The physician can then view the hash file sent by the patient. The physician prescribes the medication dose according to the patient's current condition. This information is again encrypted by the AES algorithm and stored in IPFS. Eventually, the patient can receive the physician's prescription by referring to their portal.

*c) Nurses:* The process of registering and logging into the system for the nurse is similar to physicians. Nurses' access level to patient information differs from physicians' level of access. Physicians can access all patients data, and the physicians' final approval for injecting the drug dose is done. But in emergencies where the patients health is in danger, the patient can receive permission to inject a dose of medicine by sending an emergency request to the nurse.

*d) IoT devices:* Registering IoT devices to the system is the same as registering other users. The IoT devices are mainly motion capture systems for patients with movement disorders. Motion capture systems send motion data (usually in the form of joints' angles) to the system. The data transmitted by the IoT devices are encrypted by the AES encryption algorithm and then stored as a hash file in IPFS for the physician's review or used by the decision sup port algorithm.

*e) Decision support algorithm:* The decision support algorithm is a key component of any remote management system responsible for determining the required medication dose. This algorithm is designed to automatically determine the medication dose based on the data received from the patient as well as the patients history. The determined dose is added to the patients request file as a hashed file, and after being encrypted by the AES algorithm, it is stored in IPFS and sent to the physician. Therefore, the physician may either confirm or override the determined dose by investigating the patient's condition. The main advantage of the decision support algorithm is that if it detects that the patients current state is comparable with their previous state, it can directly suggest the last dose without waiting for the physicians approval.

IV. IMPLEMENTATION REQUIREMENTS

This section suggests the tools and requirements to implement the architecture proposed in this work. Implementing the recommended model for remote management of movement disorders generally consists of two sides: the client side and the blockchain side.

1) **The Client Side:** On the client side, a distributed application (dApp) is required to connect the users to

the blockchain network. Considering the increasing popularity of mobile devices, the dApp could be installed as a smartphone application. This application could potentially facilitate the use of the system on the client (patient) side. Therefore, HTML, CSS, and Javascript seem to be necessary tools to develop the user interface.

2) **Blockchain Side:** Developing and deploying smart contracts is required on the blockchain side. Smart contracts manage data sharing and allow to implement the system functions such as sending and receiving data and authentication. Ethereum seems to be the first option for developing smart contracts on blockchain-based systems. Solidity is a high-level object-oriented language for implementing smart contracts within Ethereum. It is worth mentioning that in the first step, Remix IDE can be used as an online compiler for developing Etherium-based smart contracts.

After developing and deploying a smart contract, Node JS is used to connect Ethereum with the IPFS. As mentioned earlier, various functions must be included in smart contracts to allow the User to communicate with the IPFS. According to Fig. 2, patients, physicians, and nurses form the Users set. Functions relevant to the Users set include:

*a) User Registration:* The users initially send a registration request to the system. At first, the system checks whether the requesting User is already registered in the system or not. The Users request will be rejected if the User is already registered on the system. If a new user sends a registration request, after authentication, the new User will be registered in the system by the admin. After that, public and private keys are assigned to the registered User as the identity of system users Algorithm. 1.

---
**Algorithm 1** Registration Algorithm
---
**Require:** Request registration user
**Ensure:** Register user
  1. Send a request as registration to the system
  **if** user authentication is valid **then**
    1. admin generates the public and private key and sends them to the user
    2. push the user to the system
  **else**
    print "User is not authenticated"
  **end if**
---

*b) Upload and send files to IPFS:* Only authenticated users registered with the system can upload files. Therefore, a physician, patient, or nurse as a user sends data in the form of a file. Then the information is encrypted by the AES algorithm and stored as a hash file in the IPFS Algorithm. 2.

*c) Access to files stored in IPFS:* To access and download files from IPFS, the User must first be authenticated so that the system can ensure that the User is valid. Then the User sends the request to access and download the file from IPFS to the system. Meanwhile, the blockchain fetches the requested file from IPFS and sends it to the user Algorithm. 3.

---
**Algorithm 2** Upload File
---
**Require:** File hash
**Ensure:** Upload file
  **if** user authentication is valid **then**
    1. calculate hash file
    2. store hash file on smart contract
    3. encrypt the file using the AES algorithm
    4. push the encrypted file on the IPFS
  **else**
    print "User is not authenticated"
  **end if**
---

---
**Algorithm 3** Access file
---
**Require:** request to access file
**Ensure:** access file content
  the user request to share or access a file
  **if** user authentication is valid **then**
    1. fetch the file from the IPFS
    2. the user decrypted file and access file
    4. push encrypted file on the IPFS
  **else**
    print "User is not valid and cannot access file"
  **end if**
---

*d) Data integrity:* According to what is shown in the figures (physician, patient, or nurse), when a user stores a file in IPFS, the file hash is stored in the smart contract. On the other hand, when users request access to files stored in IPFS, after receiving the file from IPFS, the smart contract recalculates its hash file and compares it with the previous hash file. The equality of both hash files indicates the medical data integrity stored in IPFS Algorithm. 4.

---
**Algorithm 4** Data integrity
---
**Require:** request data integrity
**Ensure:** check data integrity
  1. user sends a request to check Integrity
  2. Blockchain returns previous file hash
  3. smart contract recalculates file hash as new file hash
  **if** new file hash is equal to previous file hash **then**
    1. print Integrity completed
    2. return true
  **else**
    print "Integrity does not complete"
  **end if**
---

After implementing the blockchain-based remote management system for the PD patients on the IoT platform, the PD patients managed from home should also have some necessary equipment to communicate with the proposed system. Therefore, the PD patients should be equipped with smartphones, the Internet, and motion sensors. Considering that this work focuses on movement disorders, wearable motion recording sensors should be used on the patients body. The choice of wearable motion capture sensors should make the patient feel comfortable. In addition, motion capture sensors must be equipped with IoT to send the recorded data to the blockchain system in real-time.

## V. CONCLUSION

We proposed a blockchain-based remote management model to both monitor and administer the medication dose for patients with movement disorders, especially Parkinson's disease patients. The system will be deployed on an Internet of things platform. The proposed distributed architecture aims to ensure data integrity and secure data sharing between the patient and the physician. Due to the necessity of timely treatment of patients with movement disorders, medication dosage should be controlled remotely and securely. Therefore, unlike client-server-based systems, the patient's data is stored securely and encrypted in the IPFS system in the proposed plan. We will implement the proposed system and evaluate its performance in our future work.

## REFERENCES

[1] M. Hosseinzadeh, J. Koohpayehzadeh, A. O. Bali, P. Asghari, A. Souri, A. Mazaherinezhad, M. Bohlouli, and R. Rawassizadeh, "A diagnostic prediction model for chronic kidney disease in internet of things platform," *Multimedia Tools and Applications*, vol. 80, no. 11, pp. 16 933–16 950, 2021.

[2] A. Gatouillat, Y. Badr, B. Massot, and E. Sejdić, "Internet of medical things: A review of recent contributions dealing with cyber-physical systems in medicine," *IEEE internet of things journal*, vol. 5, no. 5, pp. 3810–3822, 2018.

[3] J. Vora, P. DevMurari, S. Tanwar, S. Tyagi, N. Kumar, and M. S. Obaidat, "Blind signatures based secured e-healthcare system," in *2018 International conference on computer, information and telecommunication systems (CITS)*. IEEE, 2018, pp. 1–5.

[4] Z. Shae and J. J. Tsai, "On the design of a blockchain platform for clinical trial and precision medicine," in *2017 IEEE 37th international conference on distributed computing systems (ICDCS)*. IEEE, 2017, pp. 1972–1980.

[5] M. Elhoseny, G. Ramírez-González, O. M. Abu-Elnasr, S. A. Shawkat, N. Arunkumar, and A. Farouk, "Secure medical data transmission model for iot-based healthcare systems," *Ieee Access*, vol. 6, pp. 20 596–20 608, 2018.

[6] S. Tanwar, K. Parekh, and R. Evans, "Blockchain-based electronic healthcare record system for healthcare 4.0 applications," *Journal of Information Security and Applications*, vol. 50, p. 102407, 2020.

[7] I. Mistry, S. Tanwar, S. Tyagi, and N. Kumar, "Blockchain for 5g-enabled iot for industrial automation: A systematic review, solutions, and challenges," *Mechanical systems and signal processing*, vol. 135, p. 106382, 2020.

[8] N. Kabra, P. Bhattacharya, S. Tanwar, and S. Tyagi, "Mudrachain: Blockchain-based framework for automated cheque clearance in financial institutions," *Future Generation Computer Systems*, vol. 102, pp. 574–587, 2020.

[9] M. U. Rahman, F. Baiardi, and L. Ricci, "Blockchain smart contract for scalable data sharing in iot: a case study of smart agriculture," in *2020 IEEE Global Conference on Artificial Intelligence and Internet of Things (GCAIoT)*. IEEE, 2020, pp. 1–7.

[10] P. Sharma, R. Jindal, and M. D. Borah, "Healthify: a blockchain-based distributed application for health care," in *Applications of blockchain in healthcare*. Springer, 2021, pp. 171–198.

[11] ——, "Blockchain technology for cloud storage: A systematic literature review," *ACM Computing Surveys (CSUR)*, vol. 53, no. 4, pp. 1–32, 2020.

[12] P. Sharma, N. R. Moparthi, S. Namasudra, V. Shanmuganathan, and C.-H. Hsu, "Blockchain-based iot architecture to secure healthcare system using identity-based encryption," *Expert Systems*, vol. 39, no. 10, p. e12915, 2022.

[13] G. Gunanidhi and R. Krishnaveni, "Improved security blockchain for iot based healthcare monitoring system," in *2022 Second International Conference on Artificial Intelligence and Smart Energy (ICAIS)*. IEEE, 2022, pp. 1244–1247.

[14] P. P. Ray, D. Dash, K. Salah, and N. Kumar, "Blockchain for iot-based healthcare: background, consensus, platforms, and use cases," *IEEE Systems Journal*, vol. 15, no. 1, pp. 85–94, 2020.

[15] N. Chauhan and R. K. Dwivedi, "A secure design of the healthcare iot system using blockchain technology," in *2022 9th International Conference on Computing for Sustainable Global Development (INDIACom)*. IEEE, 2022, pp. 704–709.

[16] B. Mallikarjuna, G. Shrivastava, and M. Sharma, "Blockchain technology: A dnn token-based approach in healthcare and covid-19 to generate extracted data," *Expert Systems*, vol. 39, no. 3, p. e12778, 2022.

[17] D. W. McKee, S. J. Clement, J. Almutairi, and J. Xu, "Survey of advances and challenges in intelligent autonomy for distributed cyber-physical systems," *CAAI Transactions on Intelligence Technology*, vol. 3, no. 2, pp. 75–82, 2018.

[18] H. Zhao, P. Bai, Y. Peng, and R. Xu, "Efficient key management scheme for health blockchain," *CAAI Transactions on Intelligence Technology*, vol. 3, no. 2, pp. 114–118, 2018.

[19] P. Sharma, R. Jindal, and M. D. Borah, "A blockchain-based secure healthcare application," in *Blockchain in Digital Healthcare*. Chapman and Hall/CRC, 2021, pp. 35–54.

[20] W. Yáñez, R. Mahmud, R. Bahsoon, Y. Zhang, and R. Buyya, "Data allocation mechanism for internet-of-things systems with blockchain," *IEEE Internet of Things Journal*, vol. 7, no. 4, pp. 3509–3522, 2020.

[21] G. Srivastava, J. Crichigno, and S. Dhar, "A light and secure healthcare blockchain for iot medical devices," in *2019 IEEE Canadian conference of electrical and computer engineering (CCECE)*. IEEE, 2019, pp. 1–5.

[22] O. Novo, "Blockchain meets iot: An architecture for scalable access management in iot," *IEEE internet of things journal*, vol. 5, no. 2, pp. 1184–1195, 2018.

[23] J. Cai, X. Huang, J. Zhang, J. Zhao, Y. Lei, D. Liu, and X. Ma, "A handshake protocol with unbalanced cost for wireless updating," *IEEE Access*, vol. 6, pp. 18 570–18 581, 2018.

[24] Q. Xia, E. B. Sifah, K. O. Asamoah, J. Gao, X. Du, and M. Guizani, "Medshare: Trust-less medical data sharing among cloud service providers via blockchain," *IEEE access*, vol. 5, pp. 14 757–14 767, 2017.

[25] H. Wang and Y. Song, "Secure cloud-based ehr system using attribute-based cryptosystem and blockchain," *Journal of medical systems*, vol. 42, no. 8, pp. 1–9, 2018.

[26] G. Yang and C. Li, "A design of blockchain-based architecture for the security of electronic health record (ehr) systems," in *2018 IEEE International conference on cloud computing technology and science (CloudCom)*. IEEE, 2018, pp. 261–265.

[27] M. A. Uddin, A. Stranieri, I. Gondal, and V. Balasubramanian, "Continuous patient monitoring with a patient centric agent: A block architecture," *IEEE Access*, vol. 6, pp. 32 700–32 726, 2018.